# DNA mold-based fabrication of continuous silver nanostructures

*Christoph Hadlich, Borja Rodriguez-Barea, Darius Pohl, Bernd Rellinghaus, Artur Erbe, Ralf Seidel\**

C. Hadlich, R. Seidel
Peter Debye Institute for Soft Matter Physics, Faculty of Physics and Earth Science, University of Leipzig, Linnéstraße 5, 04103 Leipzig, Germany
E-mail: ralf.seidel@physik.uni-leipzig.de

B. Rodriguez-Barea, A. Erbe
Institute of Ion Beam Physics and Materials Research, Helmholtz-Zentrum Dresden-Rossendorf, 01328 Dresden, Germany

D.Pohl, B. Rellinghaus
Cluster of Excellence Center for Advancing Electronics Dresden (cfaed), Dresden Center for Nanoanalysis (DCN), TUD Dresden University of Technology, 01062 Dresden, Germany

Funding: This work was supported by the Deutsche Forschungsgemeinschaft through Grant SE 1646/8-2/ER 341/19-2 to R.S and A.E. and the GRK 2767 (Project No. 451785257) to R.S.,A.E. and B.R. as well as the European Union's Horizon EU research and innovation actions (RIA, 3D-BRICKS project, GA 101099125) to R.S.

**Abstract text.**

Bottom-up fabrication of inorganic nanostructures is emerging as an alternative to classical top-down approaches, offering precise nanometer-scale control at relatively low cost and effort. In particular, DNA nanostructures provide versatile scaffolds for directly templating the growth of metal structures. Previously, a DNA mold–based method for metal nanostructure synthesis has been established that supports a modular structure design and a high control over the structure formation. So far, this method was limited to the growth of gold and palladium nanostructures.

Here, we report the successful adaptation of the DNA mold-based fabrication method to produce continuous silver nanowires. By optimizing reagent concentrations and applying gentle thermal annealing, we obtain continuous wire structures of several hundred nanometer length, overcoming limitations in anisotropic growth. Despite the strong interaction of silver ions with DNA, we can control the growth without increasing the complexity of our approach. Our structures are not oxidized yet they did not exhibit conductivity. This work demonstrates the versatility of DNA-templated metallization and opens new opportunities for constructing self-assembled hybrid nanostructures with controlled shape and composition.

1. Introduction

Over the last decades DNA nanotechnology has emerged as versatile platform for arranging a vast number of different nanomaterials.[1–3] The advantages of DNA-based structure formation are the programmability by sequence-specific base pairing, sub-nanometer spatial precision and versatile chemical functionalization possibilities. These features have been exploited to organize e.g. fluorophores, polymers, and nanoparticles. In parallel, DNA has been used as template for metallization, i.e. a direct electroless metal growth. Braun et al.[4] demonstrated the fabrication of silver wires of 100 nm diameter on a linear DNA molecule. Subsequent studies achieved growth of gold,[5–8] nickel,[9–11] palladium,[12–15] platinum,[16] copper,[17,18] cobalt,[19,20] as well as non-metallic materials such as $Fe_3O_4$[21] and $CdS$[22]. In most cases, metal cations were bound to the negatively charged DNA backbone and then chemically reduced. This caused the formation of metal clusters that continued to grow in the presence of excess reactants.[23]

Advances in DNA nanotechnology have since then enabled the fabrication of DNA structures with increasingly complex arrangements and shape designs. In particular, the DNA origami method, where a long single-stranded DNA scaffold is folded with the help of numerous short staple strands, enabled the high-yield fabrication of diverse two-dimensional and three-dimensional structures.[24–26] DNA origami nanostructures have been metallized either by fully coating the structure,[27–29] or by placing reaction sites for selective metallization.[28,30–33] Furthermore, metal–semiconductor hybrid interfaces have been realized by attaching semiconducting nanoparticles prior to a final growth step.[34–36] Mineralization of DNA nanostructures has also proven effective in stabilizing DNA nanostructures[37–39]] which also supported subsequent be coating with metal oxides.[40]

However, direct ion-nucleated growth of metal on DNA nanostructures remained challenging topic due to limited control over the nucleation leading often to inhomogeneous or discontinuous structures of rather large diameters.[1,23] As alternative, pre-placement of functionalized nanoparticles to allow a seeded-dependent electroless plating offered more control over nucleation sites.[41–46] Yet, such interfacial approaches still lacked precise spatial control during growth, frequently resulting in incomplete or irregular structures. More recently, intracavitary strategies emerged that exploited the DNA nanostructure itself as a confinement. Here, metal growth is initiated from prepositioned gold nanoparticles inside a DNA cavity. Subsequent seeded growth is restricted by the internal volume of the DNA mold, which dictates and stabilizes the final shape.[47–49] Using this "DNA mold" approach, our group has shown precise pattern control and the fabrication of both elongated[50] and complex [51] shapes through controlled multimerizations of mold monomers with specific interfaces. Gold growth within the mold can reach aspect ratios of up to seven,[52] producing micrometer-long wires with metallic conductivity.[53] However, extension of this strategy to other metals has proven challenging. Only recently, the method was adapted to produce 200 nm long palladium nanostructures grown from Pd seeds.[54]

In this study, we report the fabrication of continuous elongated silver nanostructures grown with the mold-based approach. Sun et al.[47] demonstrated already the feasibility of silver growth inside DNA cavities. Still, these structures displayed limited anisotropy and yield that is required for the coalescence of multiple silver nanoparticles into continuous complex shapes. The fabrication of silver nanostructures is of interest due to their diverse applications in biomedicine,[55,56] plasmonics,[57,58] catalysis,[59,60] and sensing.[58,61] The special properties of silver nanostructures arise from their specific physicochemical properties, the strong tunable plasmonic resonance, and the high conductivity of silver. Silver ions can interact specifically with DNA by complex formation, with preference to the nitrogen atoms of guanine or cytosine bases.[62,63] or non-specifically via electrostatic interactions. With this they can even form silver-mediated base pairs.[64–67] Silver binding can alter the conductivity of DNA[68] and may even stabilize DNA nanostructures,[69] similar to Pd.[70]. However, for the DNA mold approach, the interactions with silver ions are undesirable, as they may introduce unwanted nucleation sites and may distort the DNA template. The extent of such effects depends on multiple factors, including the choice of reducing agent, [71] the presence of capping agents,[72] and light exposure.[73,74]

We demonstrate that these challenges can be overcome by optimizing the buffer, reducing agent, and reagent concentrations. With fine-tuned conditions, we obtain silver nanowires of ~26 nm diameter highlighting the versatility of the DNA mold–based approach. While the resulting silver growth is less anisotropic and more heterogeneous compared to gold, we achieved continuous and smooth nanostructures up to micrometer length by applying slight overfilling of the cavities in combination with mild thermal annealing. Energy dispersive X-ray spectroscopy (EDS) and selected area electron diffraction (SAED) analysis showed formation of crystalline silver rather than silver oxide.

2. Results and Discussion

To develop a DNA mold-based growth of silver nanostructures we employed a previously established DNA mold system (**Figure 1**).[48,50,52] The "mold" is a hollow cuboid composed of 64 parallel DNA double helices assembled by the origami method. Its outer dimensions are 40 × 25 nm, with an internal cavity of 15 × 15 nm (**Figure S1**). Attachment sites for gold nanoparticle seeds (AuNPs) inside the cavity are formed by four single-stranded DNA anchor strands protruding inwards. To form specific elongated mold multimers, each mold end contains a pattern of repulsive and interface-specific attractive strands (Figure S1). Orthogonal interface types of different patterns allow produce multimers with defined lengths up to nine units.[50] Long mold chains are obtained by mixing molds with alternating attractive interfaces in equal stoichiometry (Figure S1). To establish a simple protocol, we used gold nanoparticles as nucleation seeds, which is a common strategy for seeded silver growth.[47] Thiolated DNA–functionalized gold nanoparticles were hybridized to the anchor strands of mold monomers followed by mold multimerization. The seed-loaded multimers were then transferred into a solution containing silver nitrate ($AgNO_3$) and a reducing agent to initiate silver deposition. Due to the high reactivity of silver ions, we avoided chloride ions in the whole preparation as well as short-wavelength light following the protocol of Sun et al.[47] Initially we tested ascorbic acid (AA, $C_6H_8O_6$) as reducing agent in the presence of either Tris(hydroxymethyl)aminomethane (TRIS, $C_4H_{11}NO_3$) or boric acid as buffers both previously used for DNA-mold based metal growth.[47,48,54] Using pentameric mold structures with a single central gold seed we obtained controlled particle growth that was partially anisotropic (**Figure S2**) with similar results for both buffer. However, as TRIS is known to undergo complex formation with silver ions,[75] we decided to employ boric acid in the following experiments for the least interference with silver growth.

To determine an optimum seed:precursor ratio at which continuous silver nanostructures are obtained, we calculated the theoretical filling ratio (TFR), representing the number of silver atoms required to fill the inner mold cavity completely. As result we estimated that ~300.000 silver atoms are required per mold to achieve a TFR of 100%. We started with 50 % TFR in chain structures to produce unconnected structures (**Figure 2a**).

We next tested different reducing agents as AA is also capable of spontaneous silver nucleation on DNA in absence of seeds.[71] Sodium citrate, which has been shown to inhibit DNA-silver interactions[71] or reducing agents that in general promote anisotropic growth (e.g. tannic acid[76]) were not suitable, as they require a chemical environment that would damage the DNA nanostructures. As alternative to AA, we tested hydroxylamine (HA, $NH_2OH$), which has been used for DNA-mediated gold growth.[48] Employing HA, we observed similar results. The silver nanoparticles were, however, slightly larger at the same conditions (**see Figure S3**). The smaller silver particles in the presence of AA suggest DNA mediated spontaneous nucleation in parallel to seeded growth.[71] Since spontaneous nucleation appeared to be reduced for HA, it was used in the following growth procedures

Although silver nanoparticles appeared fully grown within 2 minutes, the growth time was extended to 10 minutes to ensure uniform deposition and consistent structural development. UV-Vis absorbance measurements showed a maximum near to 435 nm, as expected for the plasmon resonance of isotropic silver nanoparticles or the transverse mode of elongated nanoparticles (**Figure S4**).[77] The large width of the plasmon resonance peak indicated that the nanoparticles had variable aspect ratios.[78] Extending growth times further did not increase the absorbance of the silver nanostructures which underscores that the growth process was complete. After 20 minutes the absorbance started to decrease, suggesting that the particles started to sediment.[79]

To fine-tune the reaction conditions, we varied the total concentration of all reagents while keeping a 50 % TFR and a 1:1 ratio between precursor and reducing agent fixed (Figure 2a). Surprisingly, the particle size was dependent on the seed concentration despite the same number of silver ions available per mold. At 1 nM seed concentration and above we observe almost negligible growth of our seeds. Growth experiments without seeds revealed also for HA spontaneous nucleation and growth in presence (**Figure S5a**) but also in absence of DNA (**Figure S5b)** that increased with the precursor concentration. We therefore attributed the attenuated growth at high seed concentrations to non-specific nucleation that outcompeted the seeded growth. With decreasing seed concentration, seeded silver growth became dominant (Figure 2a). While at 0.5 nM the particle size was still lower than theoretically expected, appropriate filling was obtained for 0.25 nM and 0.1 nM seeds. Further lowering the seed concentration to 0.01 nM reduced the particle sizes again presumably due to an attenuation of the growth at the corresponding low precursor concentration. To further improve the reaction conditions, we optimized the ratio between precursor and reducing agent (**Figure S6**). For this, we kept the seed concentration constant at 0.25 nM and varied the concentration of either

AgNO₃ or HA. Lowering the precursor or the reducing agent led to the expected formation of smaller particles (Figure S6). However, when AgNO₃ concentrations exceeded those of the reductant, particle sizes remained constant up to a twofold excess of precursor but decreased again at a 2.5:1 ratio (Figure S6a). In addition, satellite particles began to form along the DNA chain structures, and at a 5:1 ratio, almost no growth was observed indicative of increased spontaneous nucleation. In contrast, increasing the reducing agent did not affect the particle sizes (Figure S6b). We defined a 1:1 ratio as the optimum stoichiometry for a complete reduction of the available silver precursor and applied it in the following experiments.

After establishing suitable reaction conditions, we tried to completely fill the mold multimers to obtain nanowires. To achieve this, we systematically altered the TFR between 6.25 % and 200 % using a seed concentration of 0.25 nM (Figure 2b). Particle sizes remained consistent with TFR, indicating that seeded growth dominated under the applied conditions. At high TFRs above 50 % particles began to coalesce. However, silver growth is not as spatially restricted by the presence of the DNA mold walls as seen for gold nanoparticles.[52] This led to significant stretching of the DNA at 75% and above (Figure 2b). While most particles stayed inside the cavity, single particles showed strong transverse growth with respect to the mold axis which led to outgrowth of silver at 100 % and 200 %. The outgrowth significantly hindered the filling of the DNA chain at all positions leaving the silver structures discontinuous.

To obtain a more homogeneous and controlled growth of silver along the mold cavity and a coalescence of neighboring silver nanoparticles, we doubled the number of seeds per mold. This enabled the use of TFRs of 100 % and 200 % TFR with limited outgrowth such that a more even silver coverage was obtained (**Figure 3a**). Coalescence between neighboring nanoparticles was frequently observed and we obtained well connected nanostructures (Figure 3c). Occasionally gaps between neighboring nanoparticles were still present presumably due to residual nanoparticle heterogeneity (Figure 3a). We evaluated the average length of contiguous silver segments to determine optimal conditions (Figure 3b). At 0.25 nM seed concentration and 100 % TFR, the mean segment length was 97 nm (N=338), with a highly heterogeneous distribution. Individual small particles alternated with larger segments of fully coalesced particles. Increasing the TFR to 200 % further enhanced the heterogeneity, reducing the average length of continuous segments to only 51 nm (Figure 3b, N=638). At these conditions increased outgrowth and aggregation was obtained. At 0.1 nM seed concentration and 100 % TFR particle the growth was more homogeneous but slightly reduced compared to 0.25 nM seeds yielding particles that barely connected, preventing the formation of continuous structures. (Figure 3a). To obtain coalescing particles we increased the concentration to 125 %, yielding an average segment length of 161 nm (Figure 3b, N=142). Increasing the TFR to 200 % at a seed concentration of 0.1 nM yielded an almost tripled mean segment length of 488 nm (N=32) at higher TFR of 200 %. However, this change was accompanied by a change in morphology The nanowire diameters exceeded the mold dimensions and displayed a grainier texture, with individual particles slightly outgrowing while maintaining the overall linear shape. This provided a satisfactory filling of the mold and particle coalescence, though individual gaps between particles occasionally remained (**Figure S8**). Notably, we could observe several long structures (Figure 3b) displaying longer segments with larger and smaller diameters (**Figure S6**). The reason for this heterogeneity of the growth remained however unclear.

To increase the coalescence and the homogeneity of the nanowires further, we applied thermal annealing. Structures adsorbed to the carbon films for TEM imaging were annealed at 150, 200, 250, and 300 °C for 30 minutes, and the average lengths of connected segments were analyzed (**Figure 4**). Already at 150 °C, the silver morphology changed, and the shape was smoothed, which before appeared jagged at room temperature due to the outgrown particles (Figure 4a). Annealing at 200 °C introduced thinned-out areas barely keeping the whole structures connected. Above 200 °C we observed silver structures to be fragmented with the remaining connected particles becoming larger. Finally, at 300 °C silver particles partially appeared to have lost their original shape completely (Figure 4a). Elevated temperatures are known to increase the mobility of the metal atoms which promotes Ostwald ripening[80]. Larger particles grow at the expense of smaller particles to lower the interface energy. This results in thinned regions and droplet-like large nanoparticles as most pronounced visible at 250 and 300 °C (Figure 4a). For TFR of 100 % we observed that the mean segment length of previous 97nm remained largely unchanged when annealing at 150 °C (Figure 4b). For a TFR of 200 % we obtained a small decrease from 487 nm to 335 nm or 290 nm (N=85) when annealing at 150 °C or 200 °C (N=116), respectively (Figure 4b). To test whether a stronger growth could counteract this, we also analyzed samples with a TFR of 250 % after annealing at 150 °C, which yielded an average length of 286 nm being similar to a TFR of 200 % (Figure 4b, N=102). We therefore concluded that the most homogeneous silver wires with the longest connected segments are obtained when using a 0.1 nM seed concentration, a TFR of 200 % and a thermal annealing at 150 °C.

To confirm that our nanostructures consist of silver and were not oxidized during growth nor thermal annealing, we performed selected area electron diffraction (SAED) patterns on individual silver nanostructures (**Figure 5a**) after growth, after annealing and after annealing with a subsequent H$_2$ reduction step. The observed peaks in the resulting diffractograms where in agreement with the expected peak positions and intensities of the fcc lattice of silver but not of Ag$_2$O (**Figure 5b**).[81] We observed a mild shift of ~0.5 Å of the diffraction peak at the 200 crystal plane towards larger d-spacings both for the unannealed and annealed structures (**Table S1**), suggesting a mild lattice expansion possibly due to surface oxidation.[82,83] To test a potential surface oxidation, we reduced the adsorbed nanowire structures for 24 hours in a hydrogen gas mixture. The corresponding diffractogram of this sample showed a smaller shift of ~0.2 Å for the (200) diffraction peak, suggesting that minor surface oxidation may be relevant. Since amorphous oxygen layers are by nature not detectable by diffraction, we further assessed the oxygen content of the nanowires using energy-dispersive X-ray (EDS) mapping (Figure 5c). We observed a strong signal for the L line of silver with a high signal-to-noise ratio. In contrast, the signal of the oxygen K line was only slightly increased above the background at the edges of the silver nanoparticles suggesting a minor presence of oxygen on their surfaces.

Both, SAED as well as local EDS analysis support the nanowires consist mainly of metallic silver. However, the slight shift of the lattice parameter as well as the EDS oxygen maps suggest the presence of some oxygen on the surface of the silver nanostructures which may be in the form of silver oxide or from the organic ligands used in the growth procedure.

We finally characterized the electrical properties of 17 nanowires. Individual nanowires including unannealed and annealed samples were contacted using electron beam lithography (EBL) and I-V curves were measured. To ensure clean interfaces and minimize oxidation during EBL contacting, the DNA shell was removed using mild N$_2$ plasma treatment. Despite of this precaution, all nanowires consistently exhibited insulating behavior (**Figure 6**), which was reproducible across variations in precursor concentrations, post-annealing and the use of NiCl$_2$ for nanowire deposition (see Methods). No measurable current was observed, even under voltage sweeps from 15 mV to 5 V, in contrast to previous reports of self-organized metallic nanostructures that exhibited typically metallic conduction or thermally activated transport.[53] Given the consistent non-conductivity, nanogaps between the individual silver nanoparticles are unlikely the main cause, as their heterogeneous nature would produce greater variance in conductance. Residual ligands, though not long-chain molecules, may form an insulating layer preventing coalescence at nanoparticle contact points.[84] Additionally, the higher propensity of silver compared gold to form a native oxide could contribute to surface oxidation despite our previous analyses indicating minimal oxidation. The EBL contacting process itself may alter the silver structures.[85]

3. Conclusion

In summary, we successfully established a DNA mold-based protocol for the seeded growth of elongated silver nanostructures. To ensure a specific growth inside the molds and complete filling of the molds with metal, it was particularly important to adjust the reaction concentrations and to increase the seed loading. Despite these optimizations, the overall growth remained heterogeneous limiting the coalescence of the silver nanoparticles to lengths of several hundred nanometers, Controlled overfilling of the mold more than doubled the average length and maintained DNA-guided growth, though at the expense of a grainier morphology. Using gentle thermal annealing, the structures became better coalescent and maintained the dominant metallic silver phase. Charge transport measurements on individual nanowires exhibited however an insulating behavior for all the tested samples.

Our findings highlight both the potential and the limitations of the DNA-templated fabrication of silver nanostructures. While controlled growth of elongated structures and structural continuity are achievable, conductivity remains strongly impaired. Possibly, time and post-processing dependent surface oxidation or potentially ligand-filled grain boundaries impose insulating barriers. While it is common for DNA templated metallic nanowires to exhibit strong heterogeneous conductances,[86] silver appears to be particularly sensitive to environmental factors. These findings underscore the critical influence of surface chemistry in determining nanowire conductivity and highlight the necessity for implementing effective passivation strategies to realize reliable electrical performance in nanowire-based interconnects. Nevertheless, our approach demonstrates that DNA molds can reliably direct the synthesis of silver nanostructures. Employing different mold designs,[51] will allow to create new particle shapes with specific plasmonic properties. The use of gold nanoparticles as nucleation seeds also opens the door for the fabrication of gold-silver heterostructures. We expect that the established protocol significantly expands the possibilities of DNA-templated metallization for future developments in nanoelectronics and nanoscale plasmonics.

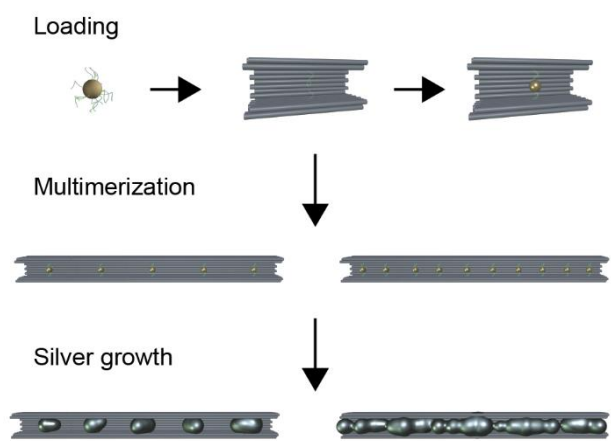

**Figure 1.** Schematic overview of the DNA mold-based silver growth. Functionalized gold nanoparticles seeds are loaded into DNA origami nanomolds by hybridization with complementary anchor strands. Depending on the number of attachment sites, one or two gold seeds are loaded per mold. In a subsequent multimerization step, molds with complimentary interfaces are mixed to form elongated structures with defined seed density. Adding silver precursor and reducing agent induces a seeded growth of silver inside the mold channels. Depending on the amount of silver grown and the seed density partially connected silver nanowire structure are obtained.

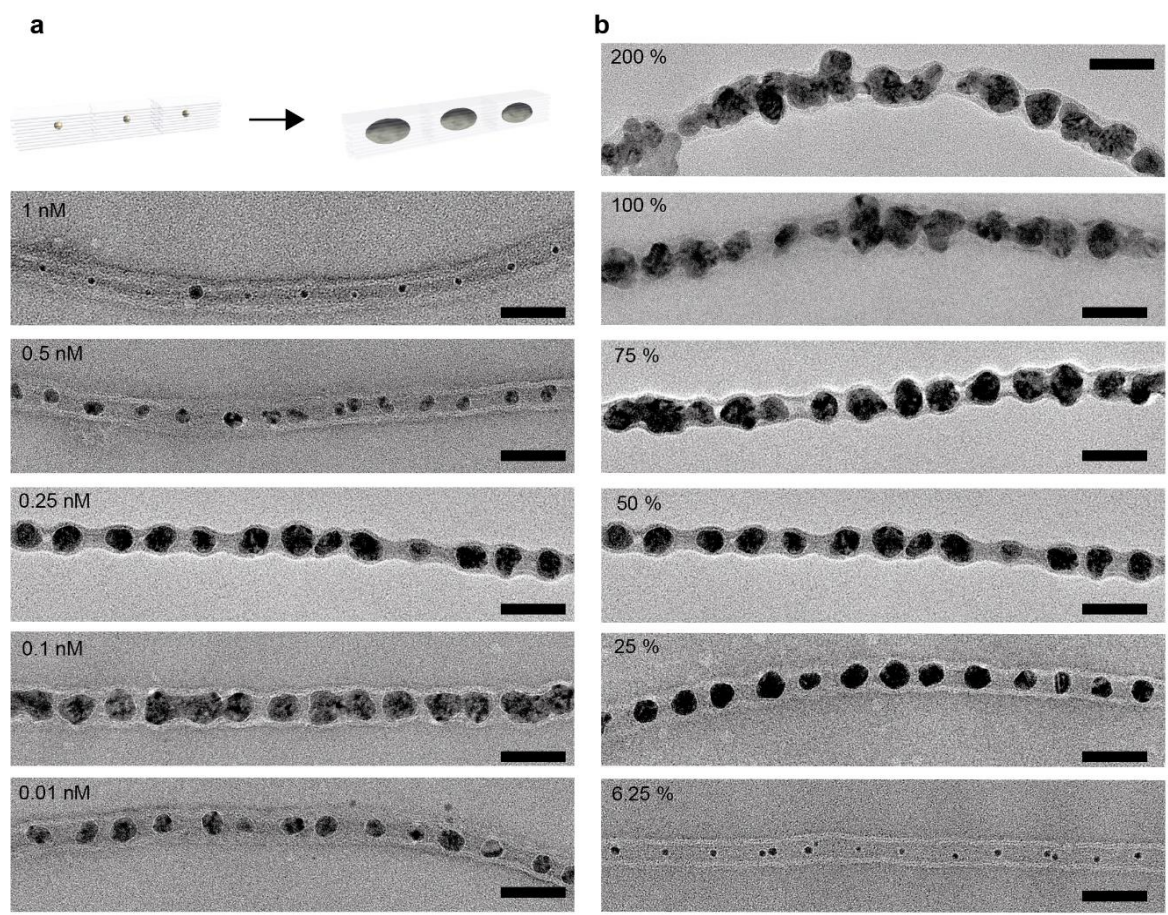

**Figure 2.** TEM images of silver nanoparticles grown inside DNA mold-chains with one seed per mold after staining. a) Silver growth was carried out at different seed (mold) concentrations as given in the images while keeping the stoichiometry of all reagents constant at a theoretical filling ratio (TFR) of 50%. b) Silver growth was carried out by varying the TFR from 6.25 % to 200 % at a fixed seed concentration of 0.25 nM. Scalebars 50 nm.

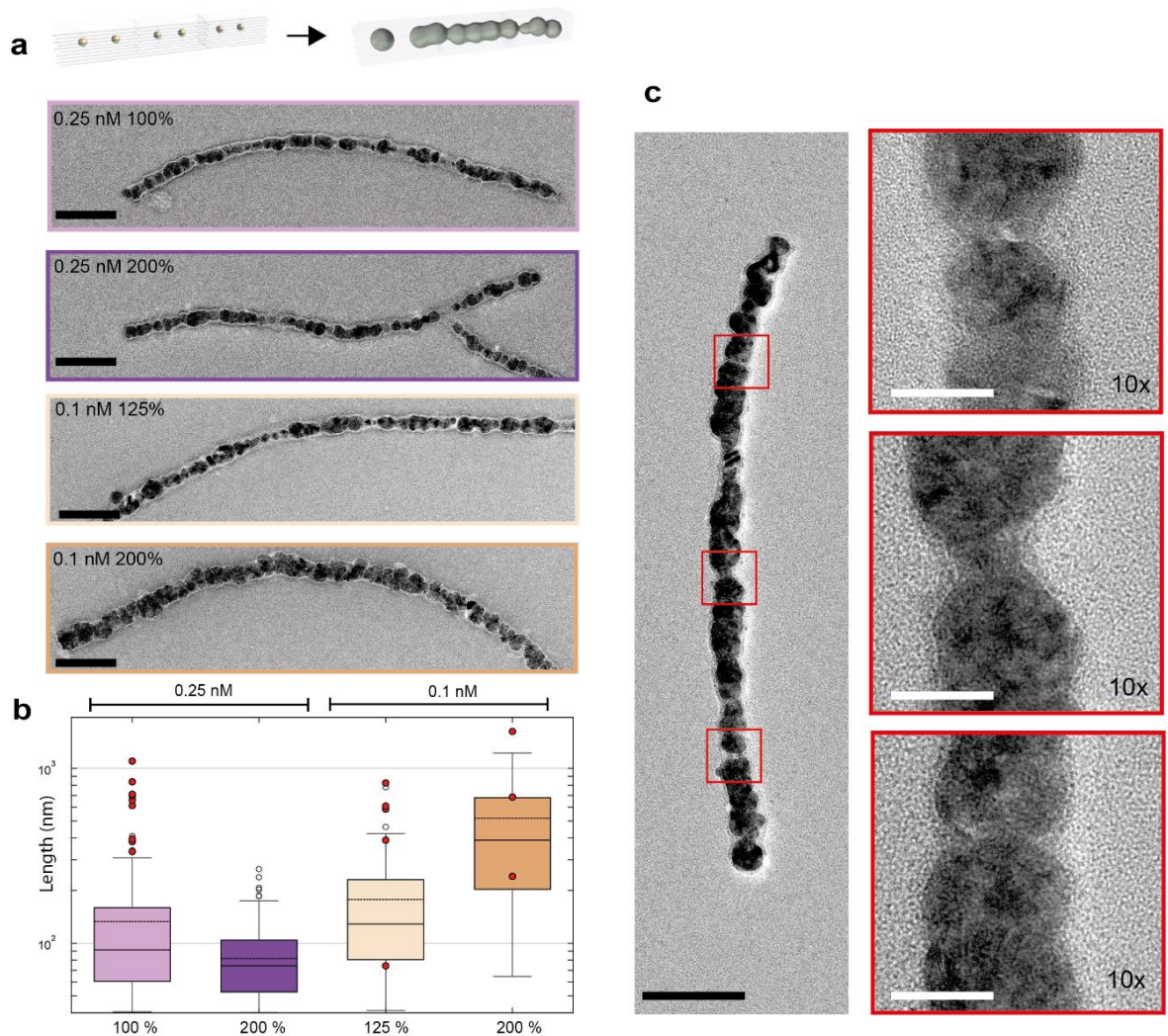

**Figure 3.** a) TEM images of silver nanostructures grown in DNA mold-chains with two seeds per mold for TFRs of 100 % (or 125%) and 200 % TFR at seed concentrations of 0.25 and 0.1 nM after staining. b) Box-and-Whisker-Plot of the length of continuous segments of the silver nanostructures for the conditions in a). The dashed lines mark the median length, the solid line the average. Boxes represent the interquartile range (25$^{th}$–75$^{th}$ percentile) with the horizontal line indicating the median; whiskers extend to the most extreme data points within 1.5× the interquartile range, and circles beyond mark outliers. Red dots mark structures which exhibited strong partial outgrowth (see Figure S7). c) TEM image of a well-connected silver nanostructure grown at 0.25 nM and a TFR of 100% in absence of staining. The high-resolution images highlight show enlarged views on the regions marked with red squares in the main image. Black scalebars are 100 nm, white scalebars 10 nm.

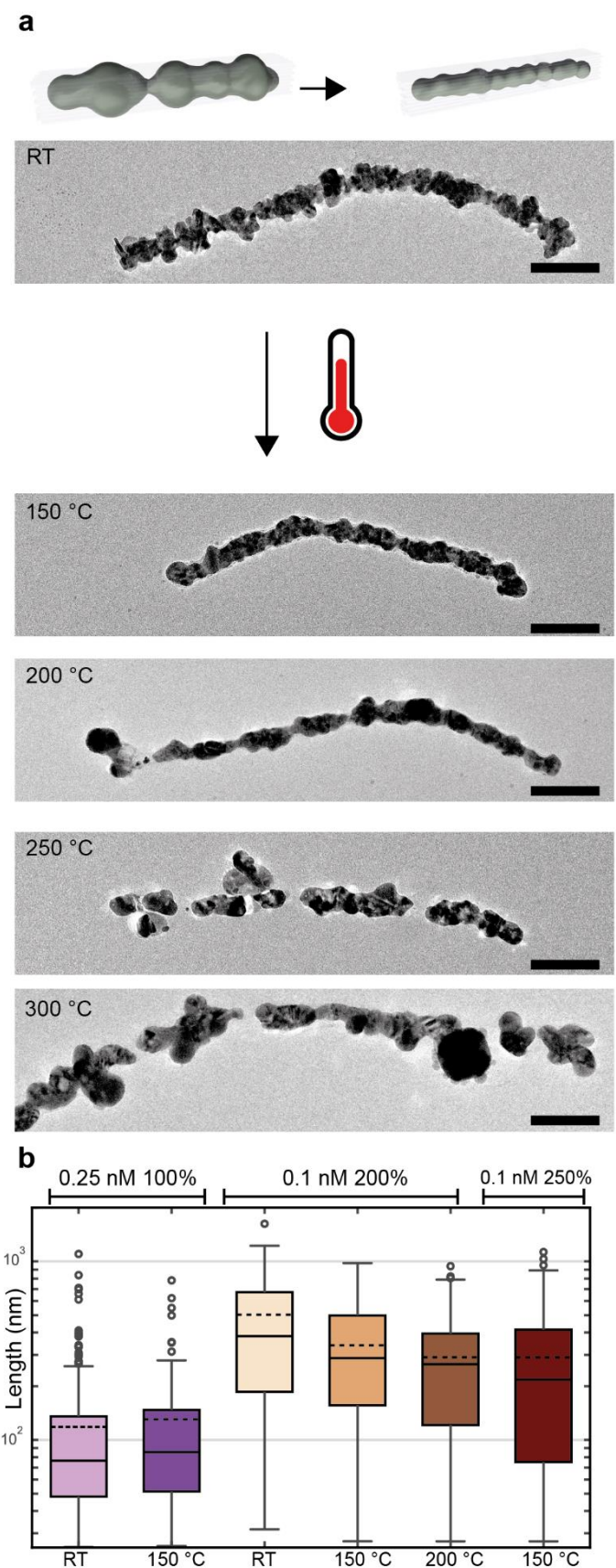

**Figure 4.** a) TEM images of silver nanostructures grown at 0.1 nM seed concentration and a TFR of 200% before and after thermal annealing for 150, 200, 250 or 300 °C. Scalebars 100

nm. b) Box-and-Whisker-Plot of the length of continuous segments of the silver nanostructures of unannealed and annealed samples. The dashed lines mark the median length, the solid line the average. Boxes represent the interquartile range (25$^{th}$–75$^{th}$ percentile) with the horizontal line indicating the median; whiskers extend to the most extreme data points within 1.5× the interquartile range, and circles beyond mark outliers.

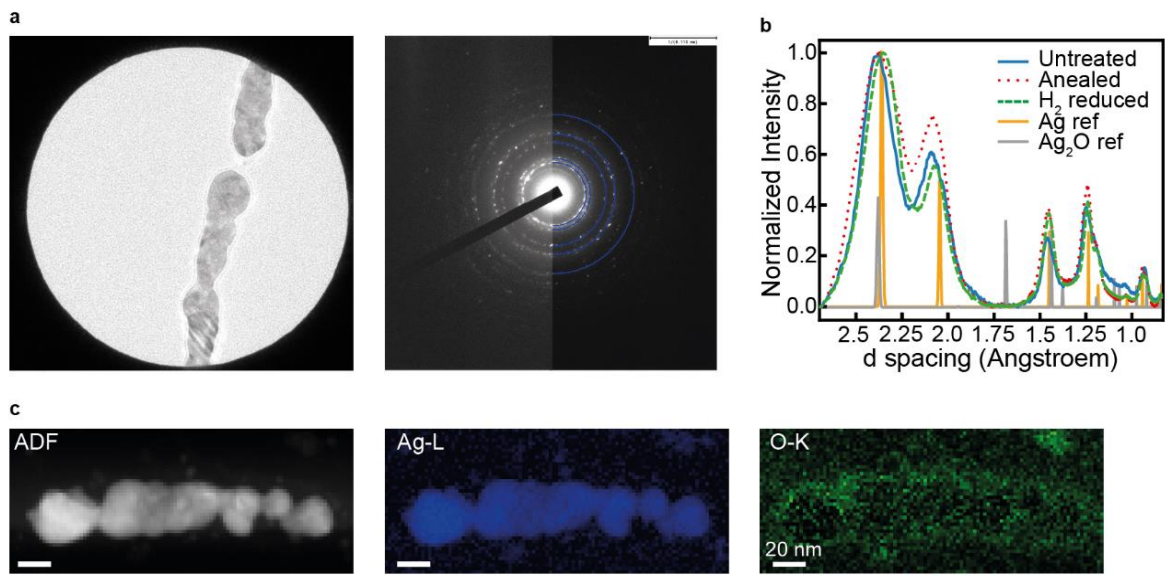

**Figure 5.** Electron diffraction and elemental analysis of the silver nanostructures. a) SAED analysis of the silver nanostructures, where electron diffraction is performed on a selected sample spot (left) for which a diffraction pattern with concentric rings is obtained (right). b) Rotation-averaged diffractograms obtained for silver nanowires after growth (blue solid line), thermal annealing at 150°C (red dotted line) and annealing with subsequent $H_2$ reduction (green dashed line). As references the expected peaks of the fcc lattices of Ag (yellow line) and $Ag_2O$ (gray line) are given.[81,87] c) EDS analysis of individual nanowires. Shown are the angular dark field (ADF) STEM image of a silver nanostructure after growth (left) and the detected X-ray intensities of the silver L line (middle) and the oxygen K line (right). Scalebars 20 nm.

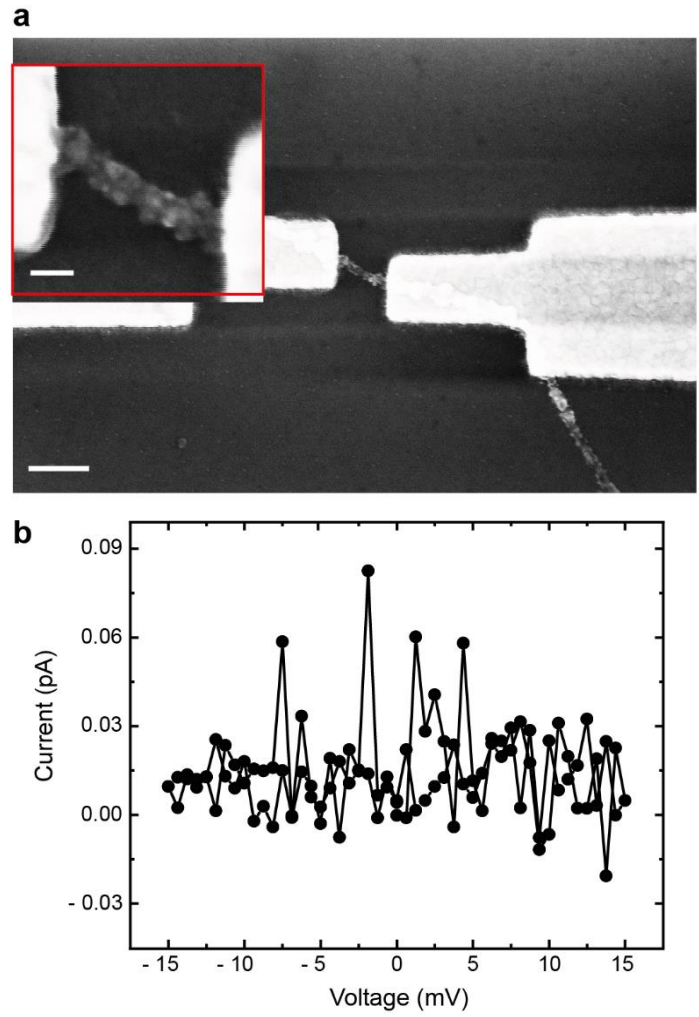

**Figure 6.** a) SEM image of a silver nanowire contacted by EBL-fabricated gold electrodes. Scalebar 200 nm in the main image, scalebar 40 nm in the inset. b) I-V curve measured for the structure shown in a) at room temperature providing no measurable current.

4. Experimental Section

*DNA origami assembly, seed loading and multimerization*: We used a previously reported DNA origami mold design including its dimerization interfaces[50]. Mold monomers were assembled in a one-pot reaction by mixing 10 nM p8064 ssDNA scaffold (Eurofins, Germany) with 100 nM staple oligonucleotides (Eurofins) including the nanoparticle capture strands. The assembly buffer consisted of 5 mM TRIS-HNO$_3$, 1 mM EDTA, 11 mM MgNO$_3$, and 5 mM NaNO$_3$ at pH 8. The mixture was first heated to 80 °C for 5 minutes, followed by slow cooling to 25 °C over 15 hours.[88] Excess staples were removed by polyethylene glycol (PEG) precipitation.[89] The concentration of DNA molds was determined by the measuring DNA absorbance at 260 nm using a NanoPhotometer P-Class P 330 (Implen, Germany). Gold nanoparticles (AuNPs) of a diameter of 5 nm (Sigma-Aldrich) were functionalized with DNA using salt-aging.[48] DNA molds were loaded with DNA-functionalized AuNP using a three-fold excess of nanoparticles per attachment site. Loading buffer, containing 0.5 mM TBE, 11 mM MgNO$_3$, and NaNO$_3$ were added yielding final concentrations of 6 nM molds, 350 mM NaNO$_3$ and 18 or 36 nM AuNPs for 1 seed/mold or 2 seed/mold, respectively. The mixture was heated to 40 °C and subsequently cooled to 23 °C at a rate of 1 K per 17 min to enable hybridization of the AuNPs to the complementary capture strands inside the mold cavity. Seed-loaded molds were purified from excess AuNPs by PEG precipitation and dissolved again in assembly buffer. For multimerization DNA mold monomers were mixed equimolar ratios in assembly buffer containing 350 mM NaNO$_3$ and incubated overnight at room temperature in the dark.

*Silver growth procedure*: The buffer solution of the seed-loaded multimers was replaced with boric buffer using PEG precipitation (0.1 M boric acid, 11 mM MgNO$_3$, pH8). All following reaction steps were done in a darkroom under illumination of a red headlamp (PETZL Arctic Core, France) and a darkroom safelight (eTONE, USA). The reducing agents ascorbic acid or hydroxylamine and silver nitrate (see result parts for final concentrations) were added in succession to boric buffer in a TC- 96well plate (F standard, Sarstedt, Germany) and mixed with the help of a magnetic stirrer (500 turns per minute). Afterwards, seed-loaded multimers were added to the mixture and the stirring continued for 1 min. The growth reaction was stopped by drop-casting the sample onto a substrate, in most cases after 10 minutes.

*TEM Imaging, SAED and EDX analysis*:

TEM imaging and SAED were carried out on a JEM2100Plus transmission electron microscope (JEOL, Japan) operated at 200 kV and equipped with a 4K CMOS camera system (TVIPS, Germany). Data acquisition was performed using the EMMenu 4.0 software package (TVIPS, Germany). EDS analysis was conducted using a, a JEOL JEM F200 STEM operated at 200 kV acceleration voltage equipped with a GATAN OneView CMOS camera for fast imaging and a dual 100 mm$^2$ window-less silicon drift detector for EDS acquisition. The EDS spectra were denoised with principal component analyses (PCA) using 3 components[90]

For sample preparation, 5 µL of the solution was applied to a plasma-cleaned, formvar/carbon-supported copper grids (PlanoEM, Germany) and incubated for 5 min. Grids were then either washed with 5 µL of HPLC water or subjected to negative staining. For negative staining, samples were briefly rinsed in three steps: one 5 µL drop of a aqueous 2% solution of uranyl formate activated with 5 mM NaOH being immediately removed after placment, a second 5 µL drop of staining solution incubated for 10s and final 5 µL drop of HPLC water for washing.

*Annealing and H$_2$ Reduction*:

Annealing and H$_2$ reduction of silver nanostructures were based on a previously reported protocol.[54] Silver structures placed on TEM-grids were plasma cleaned and then incubated in an oven for 30 min (see Results for temperatures). For the 24h H$_2$ reduction, samples on TEM-grids were placed in a flow-through chamber under a constant flow of 50 ml per minute of 10 % H$_2$ in N$_2$. During this time the chamber was heated to 150 °C.

*Contacting and conductivity measurements*: For the electrical characterization electrodes were fabricated on Si/SiO$_x$ substrates by electron beam lithography (Raith e-line Plus). The starting material consisted of p-type Si (100) wafers with a 280 nm thermally grown oxide acting as insulator. Prior to resist deposition, the substrates were cleaned in an ultrasonic bath (10 min acetone, followed by 2 min isopropanol). A bilayer resist stack of EL11 and PMMA-A4 was spin-coated and baked at 150 °C for 10 min. For e-beam exposure, a 10 kV acceleration voltage was applied, using a 120 µm aperture for defining the contact pads and a 30 µm aperture for alignment markers. Resist development was performed in an IPA/DI water mixture (7:3) for 30 s, followed by a 30 s rinse in DI. Metal electrodes were deposited using a Creavac CREAMET 600 system, starting with a 10 nm chromium adhesion layer (2 Å s$^{-1}$) and subsequently a 100 nm gold layer (5 Å s$^{-1}$). Pattern transfer was completed through an overnight lift-off step in acetone. The substrates were then exposed to O$_2$ plasma treatment to enhance the hydrophilicity and introduce negative surface charges. Subsequently, the DNA-templated silver nanowires were deposited by drop-casting and incubation for 30 min. For some samples 5 µL of 100 mM NiCl$_2$ were added at the end of the incubation period to increase surface adhesion. The prefabricated layout provided a reference framework for the localization of individual drop-cast nanowires to enable their precise electrical contacting. Nanowire positions were first determined by SEM imaging relative to the marker arrays. The coordinates were then imported into the EBL software to enable a subsequent lithography step for electrode definition directly on top of the imaged nanowires. To eliminate the organic template prior to electrode deposition, the samples underwent two $N_2$ plasma treatments (5 sccm flow, 200 W power, 25 min each; PICO, Diener Electronic–Plasma Surface Technology). Successful contacting of the nanostructures was verified by SEM. Depending on the individual nanowire length, the lithographic electrode design was adjusted, yielding a distribution of electrode–electrode spacings. Electrical characterization was performed via two-terminal I–V measurements on 17 individual nanowires under dark conditions and vacuum (base pressure $1 \times 10^{-5}$ mbar) using a Keithley 2400 source meter. The circuit was completed by positioning 25 µm tungsten probes on the gold pads. The electrical response of the nanowires was probed by performing I–V measurements with successive voltage sweeps of 15 mV, 30 mV, 3 V, and 5 V. At each applied bias, the current was carefully monitored, but no measurable current was detected, indicating the absence of conduction.

**Acknowledgements**

We would like to express our sincere thanks to Ulrich Kemper (Peter Debye Institute for Soft Matter Physics, University of Leipzig) for inital training in the laboratory as well as sharing the code for the SAED analysis. For access to and explanation of the H2 reduction funnel we gratefully thank Micheal Göpel and Hillary Msuya (Institute of Technical Chemistry, University of Leipzig). We would also like to thank David Poppitz (Institute of Technical Chemistry, University of Leipzig) for providing access to and instruction in TEM imaging, and Markus Löffler (DCN, TU Dresden) for his assistance with electron imaging.

**Data Availability Statement**

The data that support the findings of this study are available from the corresponding author upon reasonable request.

**Supporting Information**

Supporting Information is available from the Wiley Online Library or from the author.

Supporting Information

**DNA mold-based fabrication of continuous silver nanostructures**

Christoph Hadlich, Borja Rodriguez-Barea, Darius Pohl, Bernd Rellinghaus, Artur Erbe, Ralf Seidel*

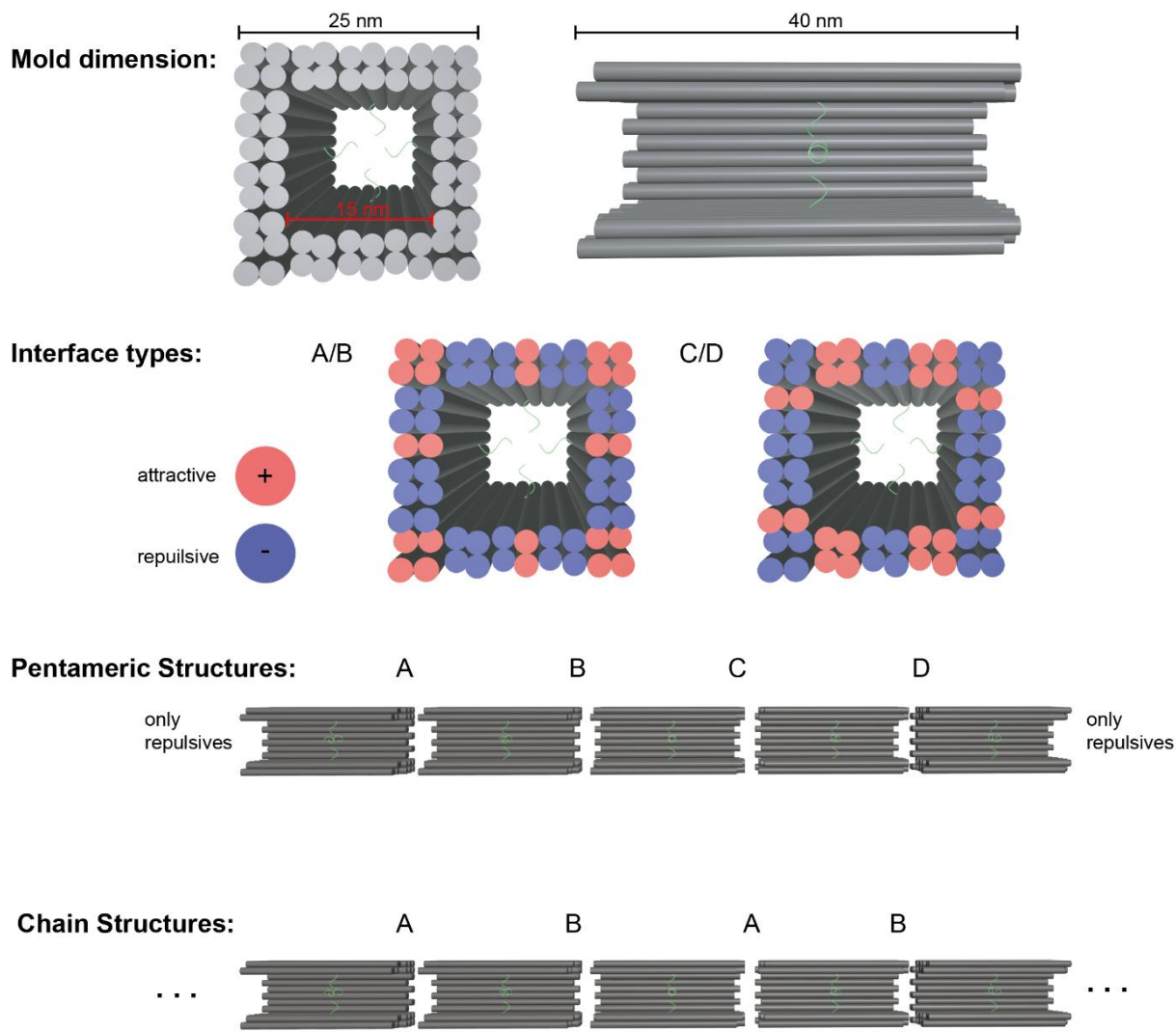

**Figure S1.** Schematic overview over the DNA mold structure and the different interface types used in this work based on Ye et al.[50] The mold walls consist of a double layer of DNA double helices. Its outer dimensions are 40 nm x 25 nm and the width of the inner cavity is 15 nm (red). Four anchor strands in the mold center (green) form a AuNP seed binding site. Helices are partially shifted so that they form a specific complementary pattern of protruding and retracted helices on the left and right side. In addition, each helix contains attractive (red) or repulsive(blue) single-stranded overhangs at both ends by which the mold has specific interfaces for multimerization[50]. There are four types of interfaces: A, B,C and D. A/B

respectively C/D share the same distribution of attractive and repulsive overhangs. However, the extension of the attractive overhangs is either on the 3′ or 5′ end thereby forming four orthogonal interfaces. Moreover, to avoid unspecific interactions molds on the ends of multimeric structures contain interfaces containing only repulsive overhangs. To form the pentameric structure we mixed at equal stoichiometry three different molds containing the interfaces: repulsive left and A right; A left and B right, B left and C right, C left and D right and D left and repulsive right. To form chains, we used two monomers with the interfaces A left and B right as well as B left and A right without any terminal repulsive interfaces.

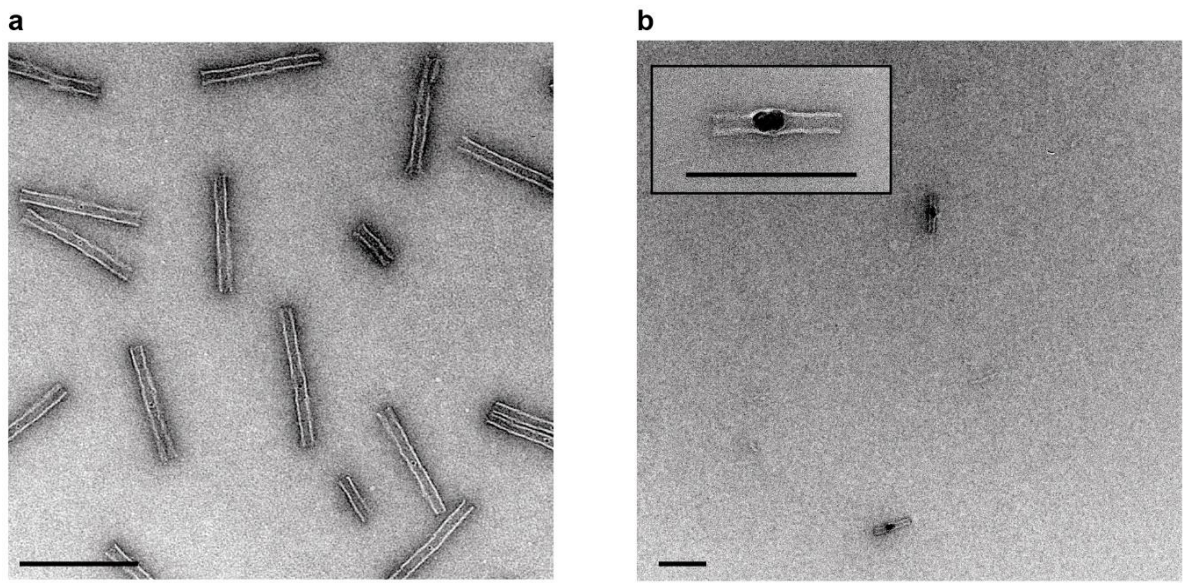

**Figure S2.** Stained TEM images of first silver growth tests using a a) Empty pentamer consisting of five DNA mold monomers and one gold nanoparticle in the central mold b) Silver growth experiment inside a pentameric structure with one central seed in the presence of Boric buffer. The inset shows shows that silver growth is well restricted by the DNA mold walls. All scalebars 200 nm.

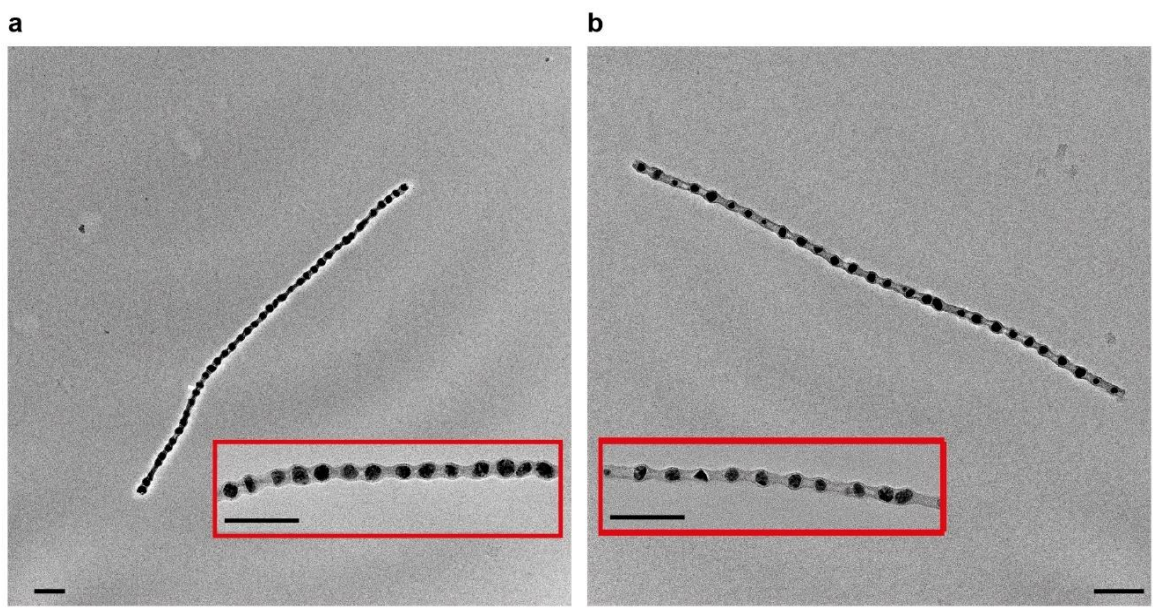

**Figure S3.** TEM images of seeded silver growth in DNA mold chains using (a) hydroxylamine (HA) and (b) ascorbic acid (AA). In the reactions 0.25 nM seeds, a TFR of 50% and 1:1 ratio between precursor and reducing agent were employed. Mean lengths of the nanoparticles along the chain axis were 22 nm (N=28) for HA and 17 nm (N=28) for AA. Scalebars 100 nm.

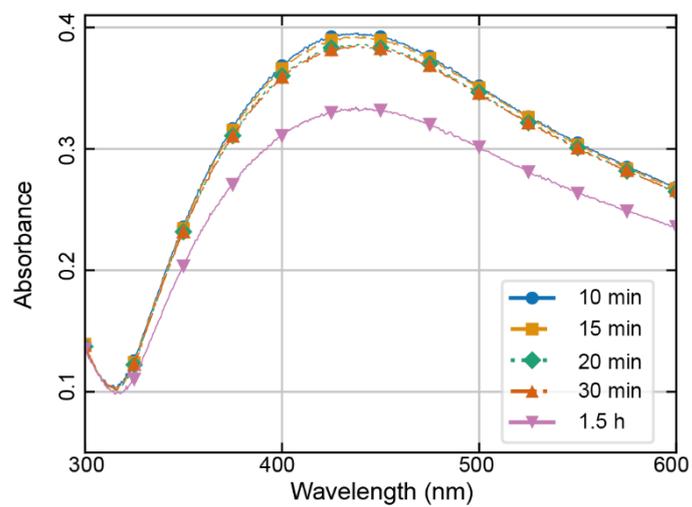

**Figure S4.** Absorbance spectra of silver nanostructures grown at 0.25 nM seed concentration, 50% TFR and a 1:1 ratio between silver precursor and hydroxyl amine. Spectra were measured at different times after seed injection.

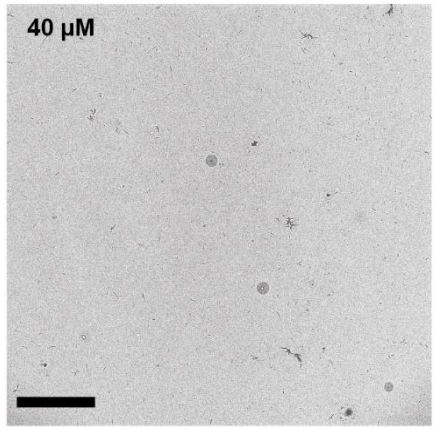
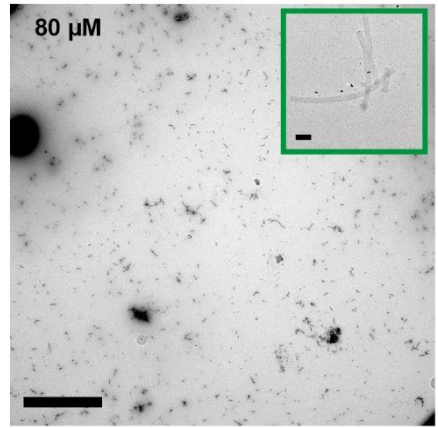
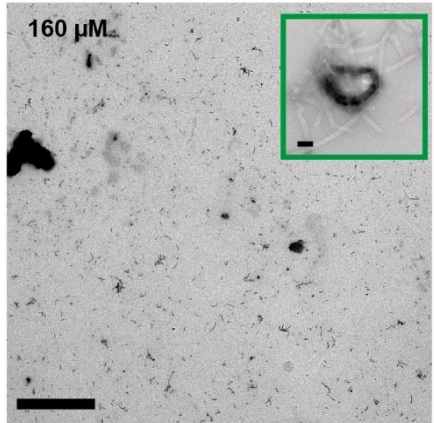
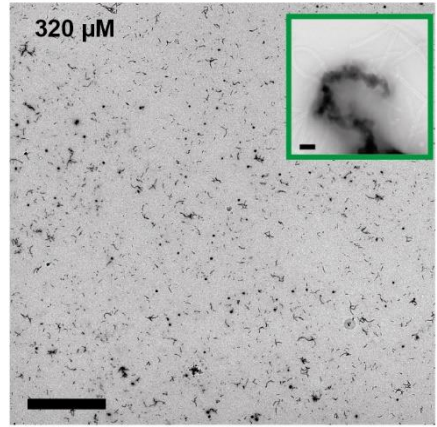
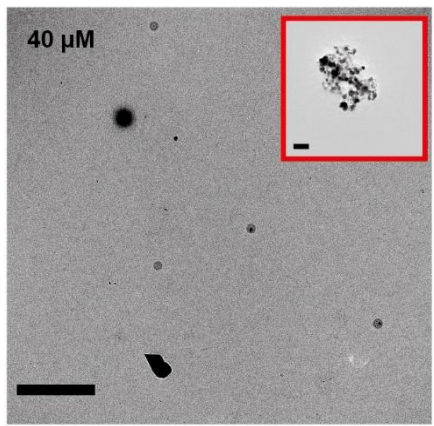
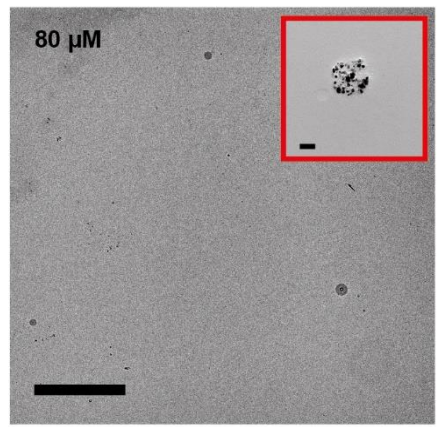
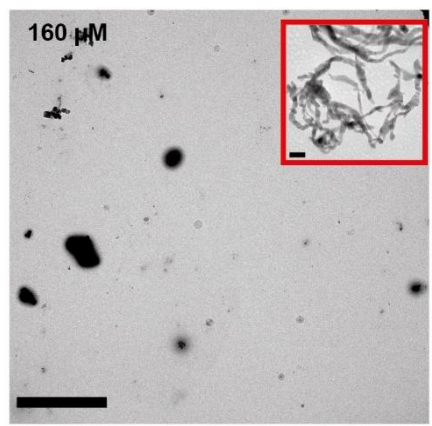
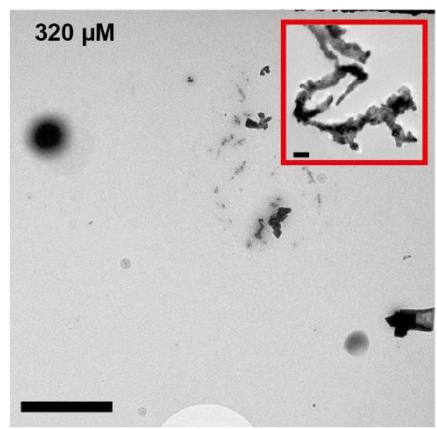

**Figure S5.** TEM overview images of silver structures grown in absence of AuNP seeds using 40 µM, 80 µM, 160 µM and 320 µM $AgNO_3$ in the presence (a) or absence (b) of DNA molds chains. Insets show enlarged views. The DNA mold concentration matched a TFR of 50 %. All samples including DNA were stained except the growth shown in the inset of the 80 µM sample. For seedless growth, we observed an increasing number of large spontaneously nucleated silver particles with increasing silver ion concentration, both with and without DNA. Without DNA we observed in general only small particle formation at 40 and 80 µM, while at higher concentrations we started to obtain larger particles in various shapes. In the presence of DNA structures, those structures were found mostly in contact with DNA causing DNA chains to aggregate. For 80 µM we could only identify a moderate interaction of silver with DNA, causing a slight staining effect and few satellite particles. For 40 µM we did not observe any DNA-relaged silver nucleation. Scalebars overview images 10 µm, scalebars insets 100 nm.

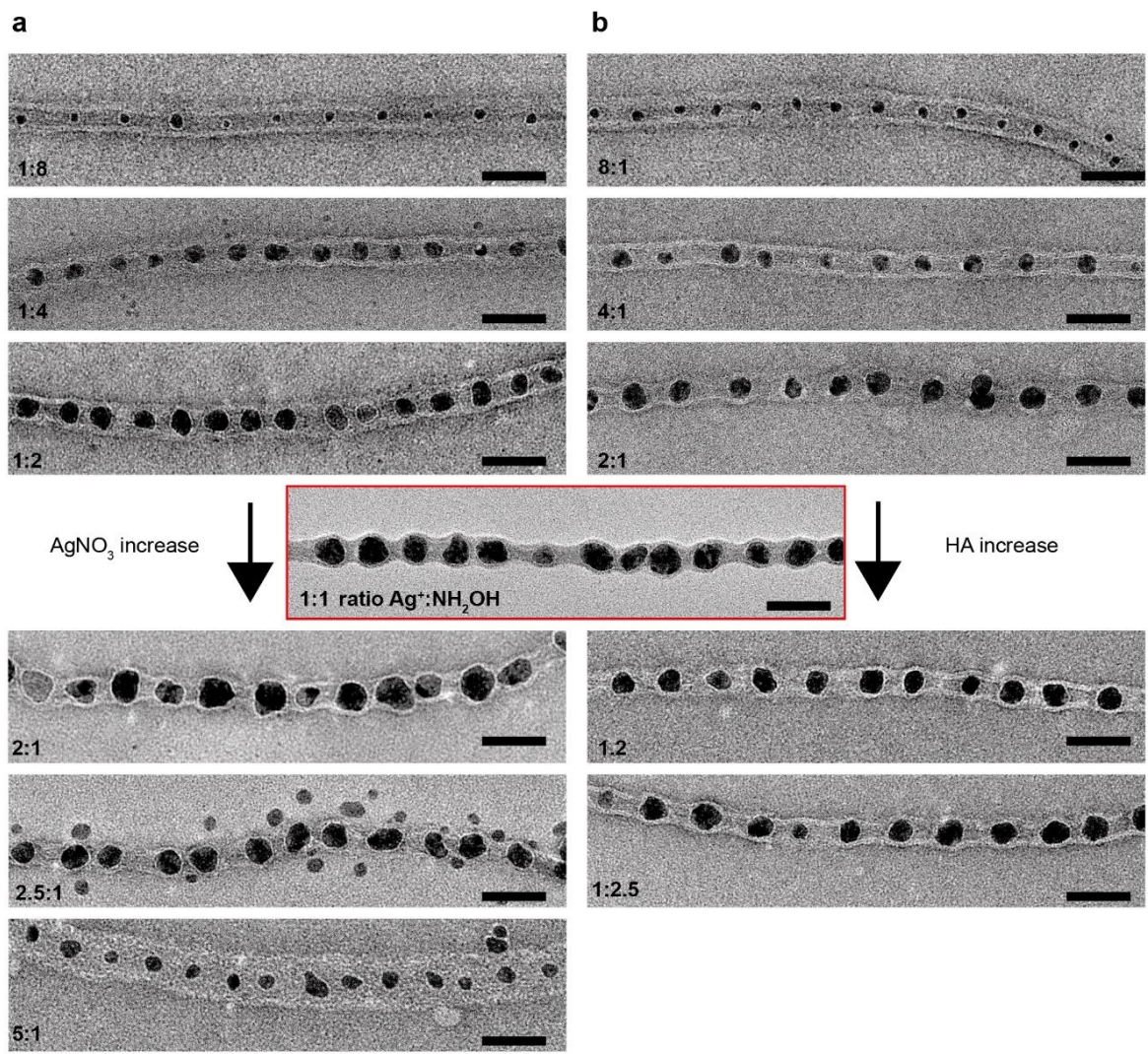

**Figure S6.** Stained TEM images of silver nanoparticles grown in DNA mold chains using different ratios of precursor / reducing agent. The seed concentration was kept at 0.25 nM. The 1:1 ratio corresponded to a 50 % TFR and thus to concentrations of silver precursor and reducing agent of 80 µM (red bordered image). a) Change of the AgNO$_3$ concentration while keeping the HA concentration constant. b) Change of the HA concentration, while keeping the the AgNO$_3$ concentration constant. Concentrations increases downwards. Scalebars 50 nm.

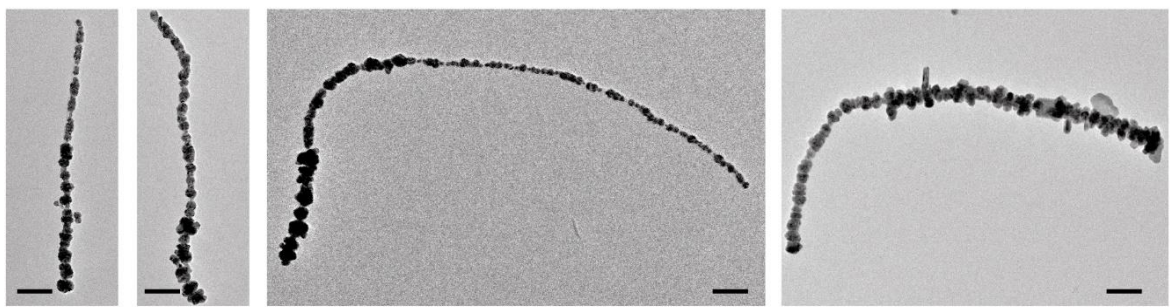

**Figure S7.** Unstained TEM images of silver nanowire structures which exhibited significant outgrowth through the mold walls. Notably, the extend of the outgrowth was different at different sections of the linear structure. Structures were grown at 0.25 nM seed concentration using 2 seeds/mold and a TFR of 100%. Scalebars 100 nm.

**Table S1.** d-spacings for the main peaks of the SAED measurements in Å

| Ag Crystal lattice planes | Ag reference[81] | Ag$_2$O reference[87] | As grown | Annealed | H$_2$ reduced |
|---|---|---|---|---|---|
| | | 2.75 | | | |
| 111 | 2.36 | 2.38 | 2.38 | 2.37 | 2.34 |
| 200 | 2.04 | 1.68 | 2.09 | 2.09 | 2.07 |
| 220 | 1.44 | 1.43 | 1.46 | 1.45 | 1.44 |
| 311 | 1.23 | 1.37 | 1.25 | 1.24 | 1.23 |
| 222 | 1.18 | 1,09 | | | |
| 331 | 0.91 | 0.97 | 0.93 | 0.94 | 0.93 |

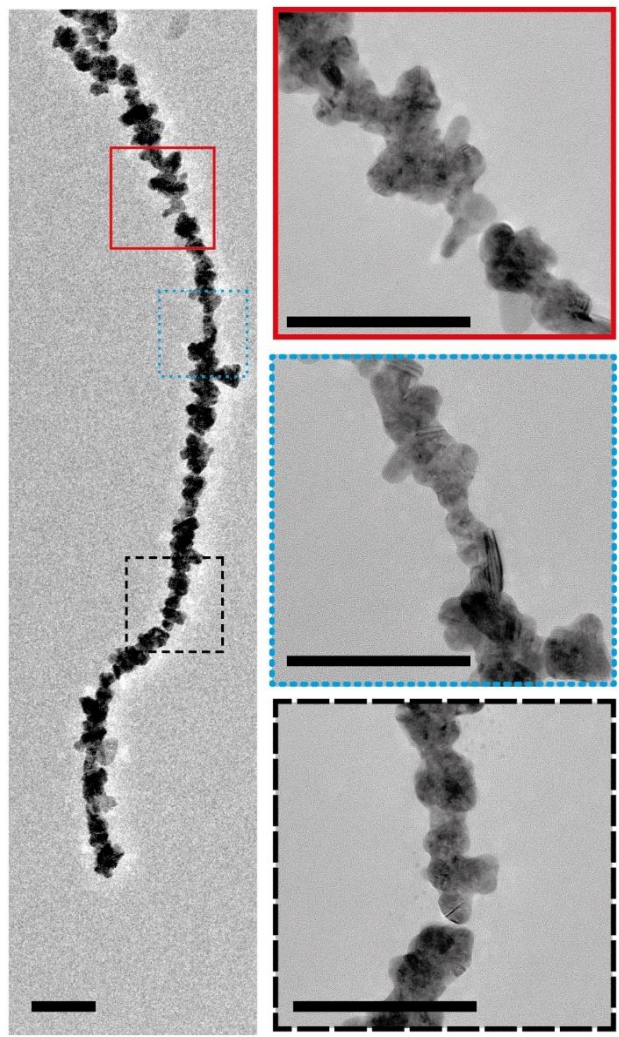

**Figure S8.** Unstained TEM image of a nanowire structure grown at 0.1 nM seeds and TFR of 200% exhibiting a grainy morphology. Insets show enlarged views on the structure revealing separated (black dashed outline), well-connected (blue dotted outline) and potentially separated (red solid) silver grains. All scalebars are 100 nm.